\documentclass[onecolumn,unpublished,a4paper]{quantumarticle}
\usepackage[utf8]{inputenc}
\usepackage[backend=biber,style=alphabetic,maxbibnames=999,backref=true]{biblatex}
\usepackage{amsmath}
\usepackage{amssymb}
\usepackage{amsthm}
\usepackage{amsfonts}
\usepackage{changepage}
\usepackage[caption=false]{subfig}
\usepackage[colorlinks]{hyperref}
\usepackage[all]{hypcap}
\usepackage{import}
\usepackage{tikz}
\usepackage{relsize}
\usepackage{color,soul}
\usepackage{capt-of}
\usepackage{mathtools}
\usepackage{float}
\usepackage[section]{placeins}
\usepackage{mdframed}
\usepackage{listings}
\usepackage[T1]{fontenc}  
\usetikzlibrary{decorations.pathreplacing}
\usepackage{braket}

\DefineBibliographyStrings{english}{
  backrefpage  = {Cited on page},
  backrefpages = {Cited on pages}
}
\renewcommand\thesection{\arabic{section}}

\theoremstyle{definition}
\newtheorem{definition}{Definition}[section]
\theoremstyle{definition}
\newtheorem{theorem}[definition]{Theorem}
\theoremstyle{definition}
\newtheorem{lemma}[definition]{Lemma}
\renewcommand{\eqref}[1]{\hyperref[#1]{Equation~\ref*{#1}}}
\newcommand{\secref}[1]{\hyperref[#1]{Section~\ref*{#1}}}
\DeclareRobustCommand{\appref}[1]{\hyperref[#1]{Appendix~\ref*{#1}}}
\newcommand{\figref}[1]{\hyperref[#1]{Figure~\ref*{#1}}}
\newcommand{\tblref}[1]{\hyperref[#1]{Table~\ref*{#1}}}
\newcommand{\theoremref}[1]{\hyperref[theorem:#1]{Theorem~\ref*{theorem:#1}}}
\newcommand{\definitionref}[1]{\hyperref[definition:#1]{Definition~\ref*{definition:#1}}}

\DeclareFixedFont{\ttb}{T1}{txtt}{bx}{n}{8}
\DeclareFixedFont{\ttm}{T1}{txtt}{m}{n}{8}
\definecolor{deepblue}{rgb}{0,0,0.5}
\definecolor{deepred}{rgb}{0.6,0,0}
\definecolor{deepgreen}{rgb}{0,0.5,0}
\newcommand{\pythonlisting}[2]{\lstinputlisting[
    showspaces=false,
    showstringspaces=false,
    showtabs=false,
    language=Python,
    basicstyle=\fontsize{#1}{#1}\ttfamily,
    commentstyle=\fontsize{#1}{#1}\ttfamily\bfseries\color{deepgreen},
    keywordstyle=\fontsize{#1}{#1}\ttfamily\bfseries\color{deepblue},
    xleftmargin=0.5cm,
    frame=tlbr,
    framesep=0.2cm,
    framerule=1pt,
    morekeywords={self,None,assert,venting\_into,venting\_into\_new\_table,alloc\_quint,quint,z,del\_measure\_x,alloc\_phase\_gradient,qpu,QPU,del\_phase\_gradient,del\_by\_equal\_to\_const,push\_uncompute\_info,pop\_uncompute\_info,ghz\_lookup},
    keywordstyle={[4]\fontsize{#1}{#1}\ttfamily\color{deepred}},
    morekeywords={[4]Q\_residue,Q\_result,Q\_total,Q\_e,Q\_grad,Q\_target,Q\_helper,Q\_unresult,Q\_l1,Q\_l0,Q\_l,Q\_k,Q\_exponent,Q\_acc,Q\_dlog,Q\_out},
]{#2}}

\begin{document}
\title{
CSS Circuits: useful quantum circuits that can be simulated with $O(1)$ cost per gate
}

\date{\today}
\author{Craig Gidney}
\email{craiggidney@google.com}
\author{Cody Jones}
\affiliation{Google Quantum AI, Santa Barbara, California 93117, USA}

\begin{abstract}
``CSS circuits'' are quantum circuits generated by $X$, $Z$, $M_X$, $M_Z$, and CNOT gates.
Although CSS circuits are extremely restrictive, they can implement important quantum protocols like teleportation, error correction, and entanglement purification.

We present a simple CSS circuit simulator that uses $O(1)$ work per gate.
The simulator corresponds to a local hidden variable model, so we use Bell inequalities to prove the simulator can't also support Hadamard gates, CZ gates, or S gates.
We note that CSS circuits appear ideal for benchmarking quantum error correction in the presence of atom loss errors.
\end{abstract}

\section{Introduction}
\label{sec:introduction}

In 1998, Daniel Gottesman showed that quantum stabilizer circuits can be efficiently simulated by classical computers~\cite{gottesmanknill}.
This is remarkable, because many important quantum protocols can be performed by stabilizer circuits.
For example: stabilizer circuits can perform quantum teleportation, can violate Bell inequalities, and can perform quantum error correction.
In fact, because stabilizer circuits are so efficient to analyze, most quantum error correction protocols created to date are based on stabilizer circuits~\cite{ErrorCorrectionZoo}.

Stabilizer circuits do have some downsides.
Of particular interest in this paper is the fact that existing stabilizer simulation methods take $\Omega(n^2)$ time per gate~\cite{aaronson2004chp,Anders2006graphsim}, where $n$ is the number of qubits in the circuit.
An error correction circuit can easily contain hundreds of thousands of gates, span thousands of qubits, and be sampled billions of times.
The only reason that kind of analysis is tractable is because it can be determined in $O(1)$ time per gate whether or not a measurement will be \emph{flipped}~\cite{gidney2021stim}.
Flip information is usually sufficient for correcting errors, and so most error correction circuit simulations use ``flip simulation'' rather than full simulation.

There are contexts where flip simulation doesn't work.
In particular, on neutral atom machines, an important kind of noise is atom loss.
When an atom is lost, the gates targeting that atom do not occur.
The missed gates can include Clifford operations.
Flip simulations understand Pauli errors, but not Clifford errors.
Consequently, the presence of atom loss noise forces the use of a full stabilizer simulation rather than a flip simulation amd this increases simulation costs from $O(1)$ per gate to $\Omega(n^2)$ per gate.

In this paper we'll show that, for a subset of stabilizer circuits that we call ``CSS circuits'', it's still possible to do fast simulation in the presence of loss.
We call this subset CSS Circuits because the states produced by such circuits always have stabilizer generators that are pure X products and pure Z products (similar to the stabilizer generators used by CSS codes).

The paper is organized as follows.
\secref{sec:methods} describes CSS circuits, shows how to simulate them, and notes limitations on extending the simulation method to other gates.
\secref{sec:conclusion} then makes some closing remarks.

\section{Body}
\label{sec:methods}

\subsection{CSS Circuits}

A  ``CSS circuit'' is a circuit built from the following five gates: $X$, $Z$, $M_X$, $M_Z$, and CNOT.
The gates may be performed probabilistically and adaptively.
The gates are defined as follows:

$$
\begin{aligned}
\text{UnitaryMatrix}(X) &= \begin{bmatrix}0&1\\1&0\end{bmatrix}\\
\text{UnitaryMatrix}(Z) &= \begin{bmatrix}1&0\\0&-1\end{bmatrix}\\
\text{UnitaryMatrix}(\text{CNOT}) &= \begin{bmatrix}1&0&0&0\\0&1&0&0\\0&0&0&1\\0&0&1&0\end{bmatrix}\\
\text{Projectors}(M_Z) &= \Big\{
\begin{bmatrix}1&0\\0&0\end{bmatrix},
\begin{bmatrix}0&0\\0&1\end{bmatrix}
\Big\}\\
\text{Projectors}(M_X) &= \Big\{
\begin{bmatrix}0.5&0.5\\0.5&0.5\end{bmatrix},
\begin{bmatrix}0.5&-0.5\\-0.5&0.5\end{bmatrix}
\Big\}
\end{aligned}$$

A CSS circuit can of course also perform any effect achievable by composition of these gates.
For example, a CSS circuit can prepare a qubit into the $|0\rangle$ state by performing a Z-basis measurement ($M_Z$) and then adaptively applying a bit flip ($X$) if the measurement result was 1 instead of 0.
CSS circuits can similarly prepare $|+\rangle$, $|-\rangle$, and $|1\rangle$.
They can also perform swap gates (by decomposing into three CNOTs), CXSWAP gates (two CNOTs), multi-qubit X basis parity measurement (like $M_{XXX}$), and multi-qubit $Z$ basis parity measurements.
CSS circuits \emph{cannot} perform Hadamard gates ($H$), CZ gates, S gates, Toffoli  gates ($CCX$), Fredkin gates ($\text{CSWAP}$), T gates, Y basis measurements ($M_Y$), or mixed basis parity measurements (like $M_{XZ}$).

\subsection{CSS Trajectories through Degenerate CSS Circuits}

A ``CSS stabilizer'' is a stabilizer with positive sign equal to a product of $Z$ operators or else to a product of $X$ operators.
For example, $Z_2 Z_9$ is a CSS stabilizer and $X_1 X_2 X_3$ is a CSS stabilizer and so is $Z_0$, but $X_2 Z_3$ isn't and neither is $Y_1$ nor $-X_0$.

A ``CSS state'' is a stabilizer state generated by CSS stabilizers.
For example, the state $|0\rangle^{\otimes n}$ is a CSS state because it is generated by the CSS stabilizers $Z_0$, $Z_1$, $\dots$, $Z_{n-1}$.

A gate has a ``CSS trajectory'' if it can produce an output state that is a CSS state when given an input that is a CSS state.
Furthermore, if the gate is a measurement, the measurement result must be 0 for the trajectory to count as a CSS trajectory.

A ``degenerate stabilizer circuit'' is a stabilizer circuit where it is possible for all measurement results to return 0 when the input state is the state $|0\rangle^{\otimes n}$.

A ``degenerate CSS circuit'' is a CSS circuit that contains no Pauli gates.

The efficient simulability of CSS circuits will come from the fact that degenerate CSS circuits are degenerate stabilizer circuits.
We prove this by proving each gate in a degenerate CSS circuit has a CSS trajectory.

\begin{lemma}
CNOT gates have a CSS trajectory.
\end{lemma}

$\text{CNOT}_{c \rightarrow t}$ has the stabilizer flow generators $Z_c \rightarrow Z_c$, $Z_t \rightarrow Z_cZ_t$, $X_c \rightarrow X_cX_t$, and $X_t \rightarrow X_t$.
These flows turn Z terms into Z terms, and X terms into X terms, so they preserve CSS-ness.
Thus, conjugating a CSS stabilizer with a CNOT gate produces a CSS stabilizer.
Therefore CNOT gates have CSS trajectories.

\qed

\begin{lemma}
$M_Z$ gates have a CSS trajectory.
\end{lemma}

There are two cases to consider: the case where the $M_Z$ measurement operation produces a deterministic result and the case where the result is random.

If the measurement is deterministic, then measurement operator $Z_m$ must be equal to a product of some subset of the $Z$ basis CSS stabilizer generators of the input state.
Because all the generators have positive sign, and positive $Z$-basis generators are closed under multiplication, the product of the CSS generators must produce $+Z_m$ rather than $-Z_m$.
Therefore, if the measurement is deterministic, the measurement result is 0.
Also, because the measurement is determined, the state does not change and so since the input is a CSS state the output is equivalently a CSS state.

In the random case there must be at least one $X$ basis generator that anticommutes with the measurement (among the $n$ CSS stabilizer generators of the $n$-qubit input state).
Choose any one of these anticommuting generators to be the ``victim''.
Multiply the victim into the other generators that anticommute with the measurement, causing those generators to now commute with the measurement.
This preserves the fact the stabilizer generators are CSS stabilizers, because X-basis CSS stabilizers are closed under multiplication.
Now only the victim anticommutes with measurement operator.
The generators of the output state are the non-victim generators and also the measurement operator $\pm Z_m$.
The sign of the new generator is determined by the random result of the measurement.
The CSS trajectory is the one where the measurement result is 0.

In both the deterministic and random case, the output state can be a CSS state.
Therefore $M_Z$ has a CSS trajectory.
\qed

\begin{lemma}
$M_X$ gates have a CSS trajectory.
\end{lemma}

The proof that X-basis measurements have CSS trajectories is identical to the proof for Z-basis measurements, but with $X$ and $Z$ exchanged.
\qed

\begin{theorem}
    Degenerate CSS circuits are degenerate stabilizer circuits.
\end{theorem}

A degenerate CSS circuit is built from CNOT gates, $M_X$ gates, and $M_Z$ gates.
These gates have CSS trajectories, and the initial state $|0\rangle^{\otimes n}$ is a CSS state, therefore the circuit as a whole has a CSS trajectory.
CSS trajectories only produce measurement results equal to 0, therefore the all-zeroes sample is possible.
Therefore a degenerate CSS circuit is a degenerate stabilizer circuit.
\qed

\subsection{CSS Simulation}

In \cite{gidney2021stim} it is shown that the difficult part of sampling from a stabilizer circuit is getting the first sample (the ``reference sample'').
If the circuit has no probabilistic or adaptive gates, further samples can be produced by diffing against the reference sample with a flip simulation.
Concretely, the simulation strategy is:

\begin{enumerate}
\item \textbf{Boostrapping}. Acquire one reference sample of the circuit (with noise removed) via some other stabilizer simulation method. This sample can be reused each time another shot is needed. The reference sample is a catalyst, not a consumable.
\item \textbf{Tracking}. Store a ``Pauli frame'' $P$ tracking whether each qubit is bit and/or phase flipped.
\item \textbf{Unitary Conjugation}.
To apply a unitary gate $G$, conjugate $P$ by $G$.
\item \textbf{Measurement Perturbation}.
The result of a measurement operation that measures $M$ is the result of the measurement recorded in the reference sample, except it's inverted if $P$ anticommutes with $M$.
\item \textbf{Stabilizer Randomization}.
When measuring or preparing an operator $M$, multiply $M$ into $P$ with 50\% probability.
(For example, to reset a qubit to $|0\rangle$, clear its tracked Pauli operator but then randomly multiply a $Z$ back into its tracked Pauli.)
\item \textbf{Noise}.
When a Pauli product noise operator $N$ is sampled, multiply $N$ into $P$.
\end{enumerate}

The key insight is to notice that, if you are told that the circuit you will need to simulate is a degenerate stabilizer circuit, then you know that the all-zero sample is a valid reference sample.
In effect, the bootstrapping step can be skipped and measurement results are purely whether the tracked Pauli frame $P$ anticommutes with the measured operator $M$.
Because all degenerate CSS circuits are degenerate stabilizer circuits, degenerate CSS circuits can be simulated in this simplified way.
This works even if gates are adaptive or probabilistic, because the reference sample is always all-zero.
Intended Pauli gates can then be accounted for by treating them as noise, yielding a straightforward simulation strategy for all CSS circuits.

In \figref{fig:sim} we present simple Python code that implements CSS circuit simulation.
It's trivial to see that the simulation strategy performs $O(1)$ work per gate.
Each qubit $q$ is represented as two bits: one bit $x_q$ indicating if the qubit is bit flipped and one bit $z_q$ indicating if the qubit is phase flipped.
Qubits are initialized into the $|0\rangle$ state by setting $x_q$ to false and randomizing $z_q$.
The Pauli X gate flips $x_q$.
The Pauli Z gate flips $z_q$.
The $M_X$ gate reports $z_q$ while randomizing $x_q$.
The $M_Z$ gate reports $x_q$ while randomizing $z_q$.
The CNOT gate controlled by $c$ and targeting $t$ xors $x_c$ into $x_t$ and also xors $z_t$ into $z_c$.
All of these implementations apply $O(1)$ operations to $O(1)$ bits, and so the simulator performs $O(1)$ work per gate.

\begin{figure}[H]
\pythonlisting{7pt}{simulator.py}
\caption{
\label{fig:sim}
A CSS circuit simulator in twenty lines of Python.
}
\end{figure}

\subsection{CSS + Hadamard Violates Bell Inequalities}

Looking at the form of the code in \figref{fig:sim}, it seems easy to add support for Hadamard gates.
Hadamard exchanges the X and Z axes of a qubit, and so one might expect the Hadamard can be simulated by simply swapping the X and Z bits associated with a qubit.
However, this is wrong.
For example, the resulting simulation code acts as if $(H_a \cdot \text{CNOT}_{a \rightarrow b})^4$ has no effect when actually that gate sequence is equivalent to $X_b$.
In fact, there is a foundational obstacle to adding support for the Hadamard gate to this simulator.

Note that the simulator in \figref{fig:sim} corresponds to a local hidden variable model.
Each qubit is represented by two bits, and those bits directly specify the results of possible measurements of the qubit.
Bell inequalities cannot be violated by local hidden variable models~\cite{Bell1964}.
Therefore, if CSS+H circuits can violate Bell inequalities, the simulator cannot possibly support Hadamard gates without some sort of major change.

The GHZ test~\cite{ghztest} has a Bell inequality.
In this test, three spacelike separated players are given respective bits $x, y, z$.
The bits are promised to satisfy $(x + y + z) \bmod 2 = 0$.
The players produce respective bits $A, B, C$.
They pass the test if $(A + B + C) \bmod 2 = (x+y+z)/2$.
When $x, y, z$ are chosen uniformly randomly, classical players fail the test at least 25\% of the time~\cite{ghztest}.
Quantum players can pass 100\% of the time by consuming entanglement~\cite{ghztest}.

\begin{figure}[H]
    \centering
    \includegraphics[width=0.85\linewidth]{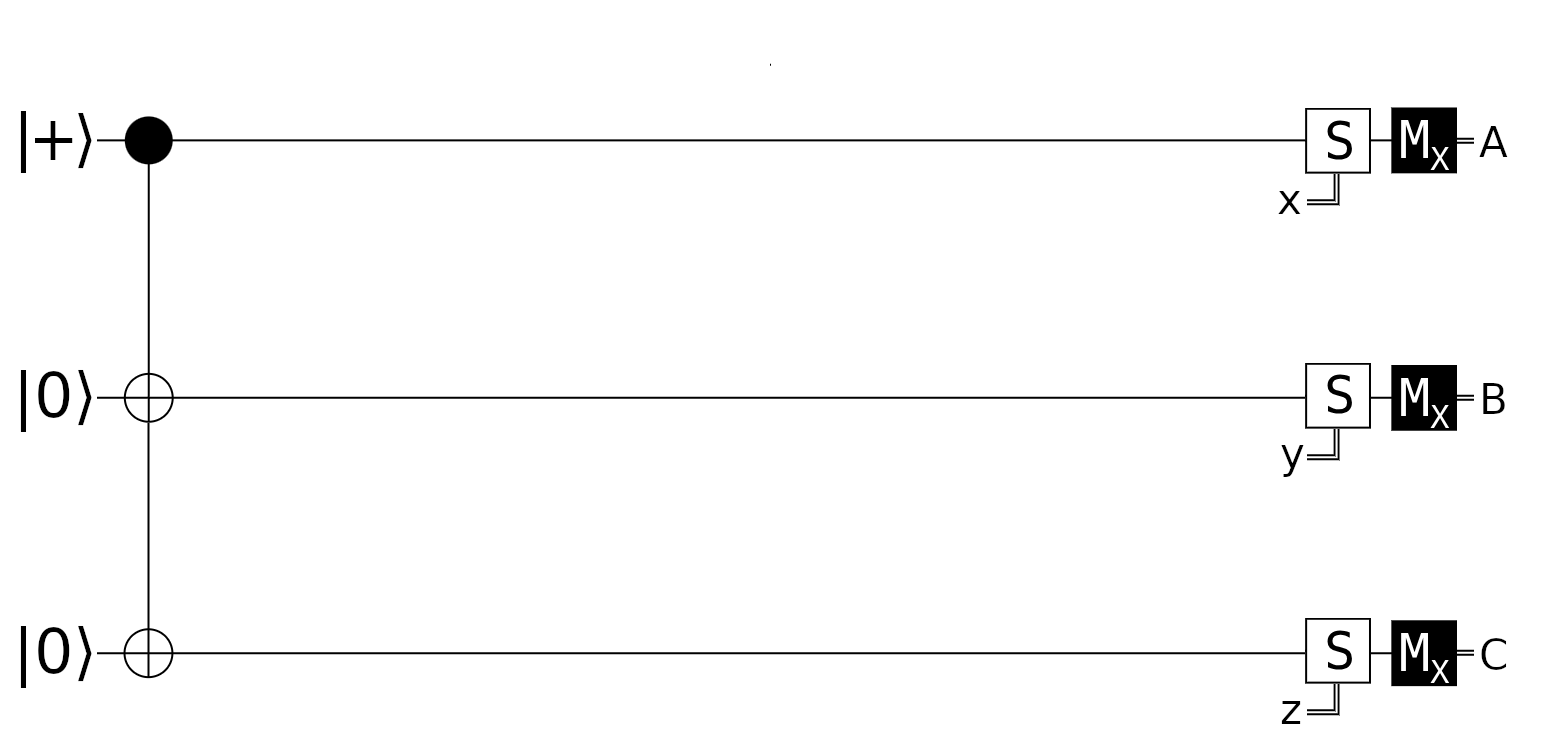}
    \caption{
        A CSS + S circuit that passes the GHZ test.
        Left half is the preparation of the entangled state before the test. Right half is the spacelike separated portion used during the test. \href{https://algassert.com/quirk\#circuit=\%7B\%22cols\%22\%3A\%5B\%5B\%22H\%22\%5D\%2C\%5B\%22\%E2\%80\%A2\%22\%2C\%22X\%22\%2C\%22X\%22\%5D\%2C\%5B1\%2C1\%2C1\%2C\%22Measure\%22\%2C\%22Measure\%22\%2C\%22Measure\%22\%2C\%22Measure\%22\%2C\%22Measure\%22\%2C\%22Measure\%22\%5D\%2C\%5B1\%2C1\%2C1\%2C\%22X\%22\%2C\%22X\%22\%2C1\%2C\%22\%E2\%80\%A2\%22\%2C\%22\%E2\%97\%A6\%22\%5D\%2C\%5B1\%2C1\%2C1\%2C1\%2C\%22X\%22\%2C\%22X\%22\%2C\%22\%E2\%97\%A6\%22\%2C\%22\%E2\%80\%A2\%22\%5D\%2C\%5B1\%2C1\%2C1\%2C\%22X\%22\%2C1\%2C\%22X\%22\%2C\%22\%E2\%80\%A2\%22\%2C\%22\%E2\%80\%A2\%22\%5D\%2C\%5B1\%2C1\%2C1\%2C\%22Chance3\%22\%5D\%2C\%5B\%22Z\%5E\%C2\%BD\%22\%2C1\%2C1\%2C\%22\%E2\%80\%A2\%22\%5D\%2C\%5B1\%2C\%22Z\%5E\%C2\%BD\%22\%2C1\%2C1\%2C\%22\%E2\%80\%A2\%22\%5D\%2C\%5B1\%2C1\%2C\%22Z\%5E\%C2\%BD\%22\%2C1\%2C1\%2C\%22\%E2\%80\%A2\%22\%5D\%2C\%5B\%22H\%22\%2C\%22H\%22\%2C\%22H\%22\%5D\%2C\%5B\%22Measure\%22\%2C\%22Measure\%22\%2C\%22Measure\%22\%5D\%2C\%5B\%22zpar\%22\%2C\%22zpar\%22\%2C\%22zpar\%22\%2C1\%2C1\%2C1\%2C1\%2C1\%2C\%22X\%22\%5D\%2C\%5B1\%2C1\%2C1\%2C1\%2C1\%2C1\%2C\%22\%E2\%97\%A6\%22\%2C\%22\%E2\%97\%A6\%22\%2C\%22X\%22\%5D\%2C\%5B1\%2C1\%2C1\%2C1\%2C1\%2C1\%2C1\%2C1\%2C\%22Chance\%22\%5D\%5D\%2C\%22init\%22\%3A\%5B0\%2C0\%2C0\%2C0\%2C0\%2C0\%2C\%22\%2B\%22\%2C\%22\%2B\%22\%5D\%7D}{Click here to open in Quirk.}
    }
    \label{fig:ghz-s}
\end{figure}

\begin{figure}[H]
    \centering
    \includegraphics[width=0.85\linewidth]{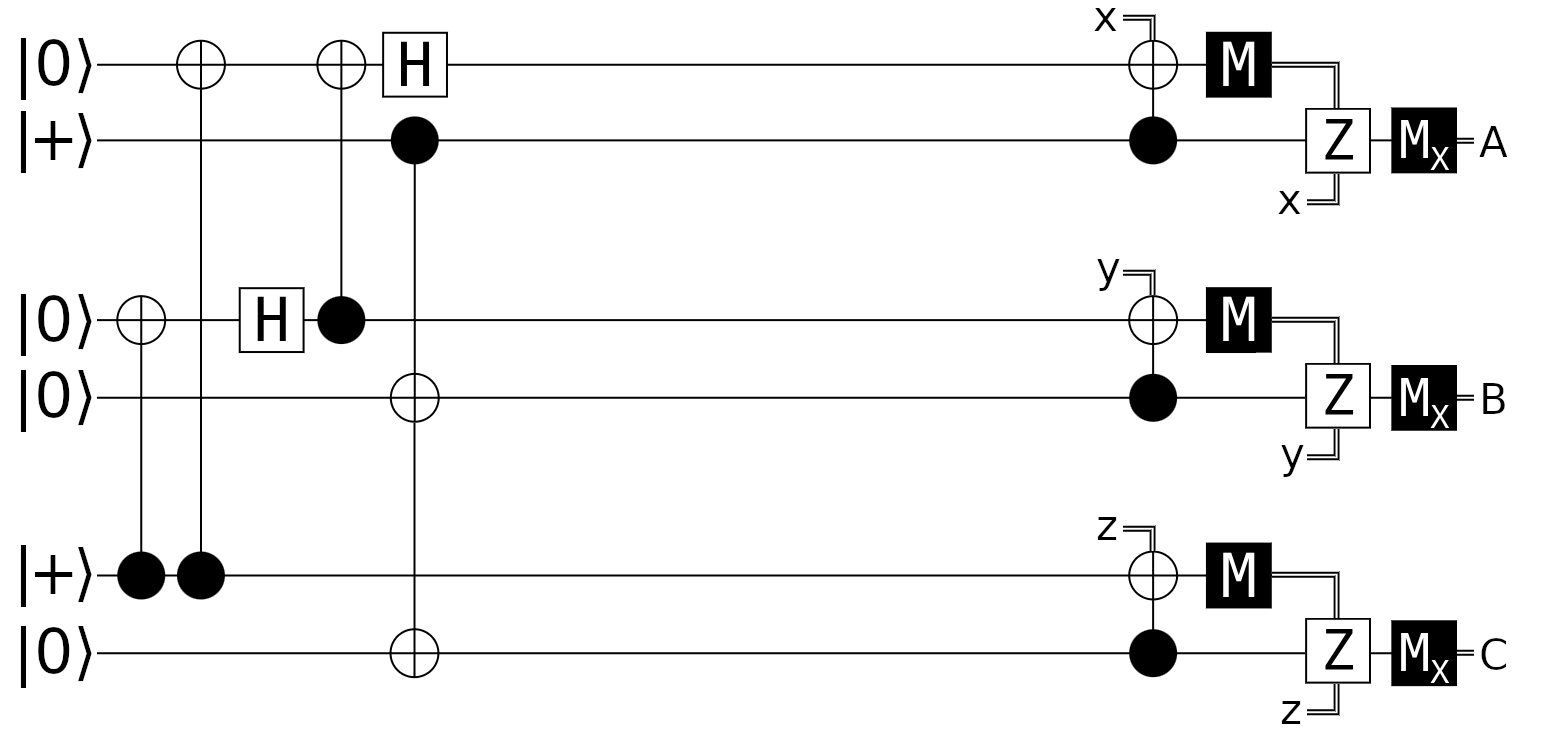}
    \caption{
        A CSS + H circuit that passes the GHZ test.
        Left half is the preparation of the entangled state before the test. Right half is the spacelike separated portion used during the test. \href{https://algassert.com/quirk\#circuit=\%7B\%22cols\%22\%3A\%5B\%5B\%22X\%22\%2C1\%2C\%22X\%22\%2C1\%2C\%22\%E2\%80\%A2\%22\%5D\%2C\%5B1\%2C1\%2C\%22H\%22\%5D\%2C\%5B\%22X\%22\%2C1\%2C\%22\%E2\%80\%A2\%22\%5D\%2C\%5B\%22H\%22\%5D\%2C\%5B1\%2C\%22\%E2\%80\%A2\%22\%2C1\%2C\%22X\%22\%2C1\%2C\%22X\%22\%5D\%2C\%5B1\%2C1\%2C1\%2C1\%2C1\%2C1\%2C\%22Measure\%22\%2C\%22Measure\%22\%2C\%22Measure\%22\%2C\%22Measure\%22\%2C\%22Measure\%22\%2C\%22Measure\%22\%5D\%2C\%5B1\%2C1\%2C1\%2C1\%2C1\%2C1\%2C\%22X\%22\%2C\%22X\%22\%2C1\%2C\%22\%E2\%80\%A2\%22\%2C\%22\%E2\%97\%A6\%22\%5D\%2C\%5B1\%2C1\%2C1\%2C1\%2C1\%2C1\%2C1\%2C\%22X\%22\%2C\%22X\%22\%2C\%22\%E2\%97\%A6\%22\%2C\%22\%E2\%80\%A2\%22\%5D\%2C\%5B1\%2C1\%2C1\%2C1\%2C1\%2C1\%2C\%22X\%22\%2C1\%2C\%22X\%22\%2C\%22\%E2\%80\%A2\%22\%2C\%22\%E2\%80\%A2\%22\%5D\%2C\%5B1\%2C1\%2C1\%2C1\%2C1\%2C1\%2C\%22Chance3\%22\%5D\%2C\%5B\%22X\%22\%2C\%22\%E2\%80\%A2\%22\%2C1\%2C1\%2C1\%2C1\%2C\%22\%E2\%80\%A2\%22\%5D\%2C\%5B1\%2C1\%2C\%22X\%22\%2C\%22\%E2\%80\%A2\%22\%2C1\%2C1\%2C1\%2C\%22\%E2\%80\%A2\%22\%5D\%2C\%5B1\%2C1\%2C1\%2C1\%2C\%22X\%22\%2C\%22\%E2\%80\%A2\%22\%2C1\%2C1\%2C\%22\%E2\%80\%A2\%22\%5D\%2C\%5B\%22Measure\%22\%2C1\%2C\%22Measure\%22\%2C1\%2C\%22Measure\%22\%5D\%2C\%5B\%22\%E2\%80\%A2\%22\%2C\%22Z\%22\%2C1\%2C1\%2C1\%2C1\%2C\%22\%E2\%80\%A2\%22\%5D\%2C\%5B1\%2C1\%2C\%22\%E2\%80\%A2\%22\%2C\%22Z\%22\%2C1\%2C1\%2C1\%2C\%22\%E2\%80\%A2\%22\%5D\%2C\%5B1\%2C1\%2C1\%2C1\%2C\%22\%E2\%80\%A2\%22\%2C\%22Z\%22\%2C1\%2C1\%2C\%22\%E2\%80\%A2\%22\%5D\%2C\%5B1\%2C\%22H\%22\%2C1\%2C\%22H\%22\%2C1\%2C\%22H\%22\%5D\%2C\%5B1\%2C\%22Measure\%22\%2C1\%2C\%22Measure\%22\%2C1\%2C\%22Measure\%22\%5D\%2C\%5B1\%2C\%22zpar\%22\%2C1\%2C\%22zpar\%22\%2C1\%2C\%22zpar\%22\%2C1\%2C1\%2C1\%2C1\%2C1\%2C\%22X\%22\%5D\%2C\%5B1\%2C1\%2C1\%2C1\%2C1\%2C1\%2C1\%2C1\%2C1\%2C\%22\%E2\%97\%A6\%22\%2C\%22\%E2\%97\%A6\%22\%2C\%22X\%22\%5D\%2C\%5B1\%2C1\%2C1\%2C1\%2C1\%2C1\%2C1\%2C1\%2C1\%2C1\%2C1\%2C\%22Chance\%22\%5D\%5D\%2C\%22init\%22\%3A\%5B0\%2C\%22\%2B\%22\%2C0\%2C0\%2C\%22\%2B\%22\%2C0\%2C0\%2C0\%2C0\%2C\%22\%2B\%22\%2C\%22\%2B\%22\%5D\%7D}{Click here to open in Quirk.}  
    }
    \label{fig:ghz-h}
\end{figure}

\begin{figure}[H]
    \centering
    \includegraphics[width=0.85\linewidth]{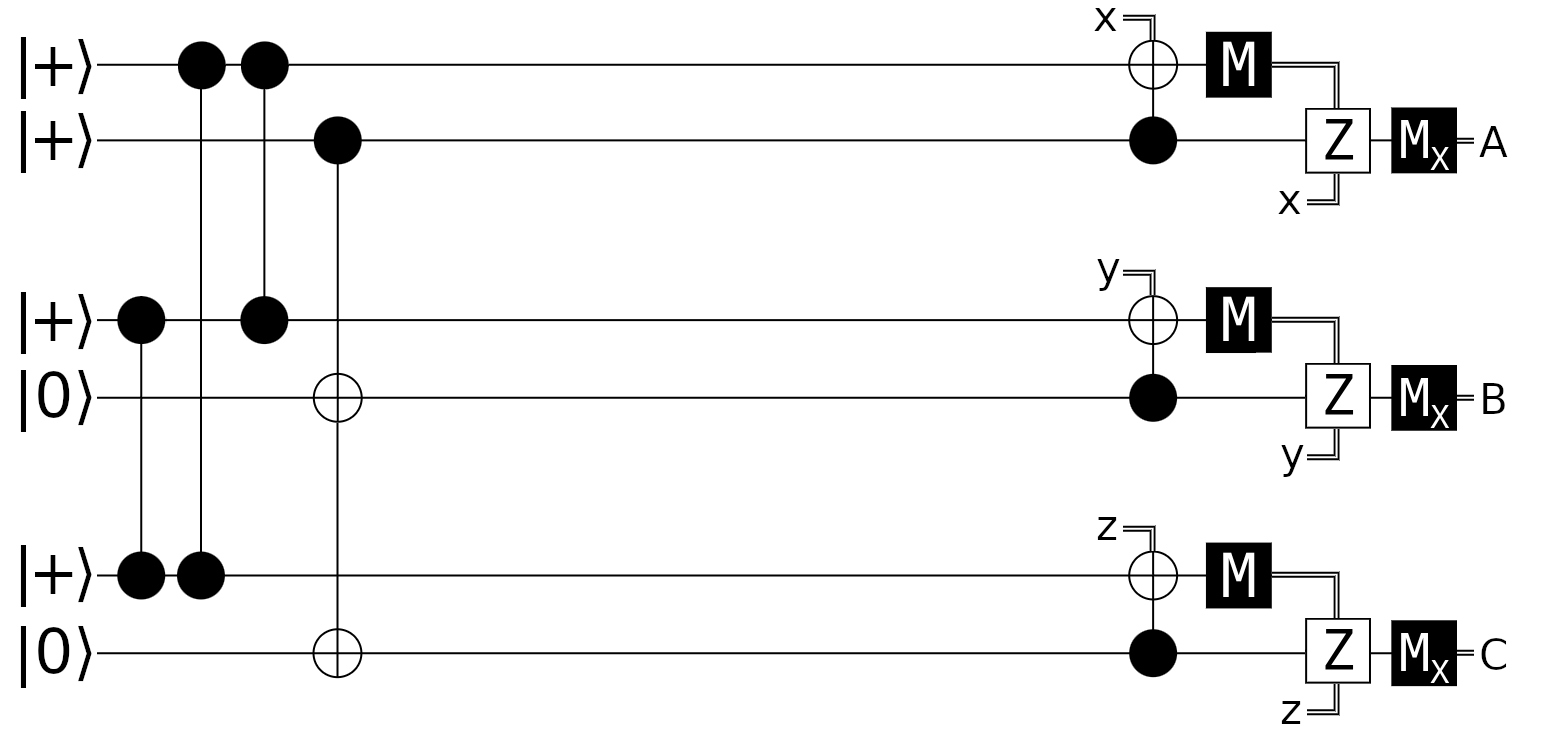}
    \caption{
        A CSS + CZ circuit that passes the GHZ test.
        Left half is the preparation of the entangled state before the test. Right half is the spacelike separated portion used during the test. \href{https://algassert.com/quirk\#circuit=\%7B\%22cols\%22\%3A\%5B\%5B\%22Z\%22\%2C1\%2C\%22Z\%22\%2C1\%2C\%22\%E2\%80\%A2\%22\%5D\%2C\%5B\%22Z\%22\%2C1\%2C\%22\%E2\%80\%A2\%22\%5D\%2C\%5B1\%2C\%22\%E2\%80\%A2\%22\%2C1\%2C\%22X\%22\%2C1\%2C\%22X\%22\%5D\%2C\%5B1\%2C1\%2C1\%2C1\%2C1\%2C1\%2C\%22Measure\%22\%2C\%22Measure\%22\%2C\%22Measure\%22\%2C\%22Measure\%22\%2C\%22Measure\%22\%2C\%22Measure\%22\%5D\%2C\%5B1\%2C1\%2C1\%2C1\%2C1\%2C1\%2C\%22X\%22\%2C\%22X\%22\%2C1\%2C\%22\%E2\%80\%A2\%22\%2C\%22\%E2\%97\%A6\%22\%5D\%2C\%5B1\%2C1\%2C1\%2C1\%2C1\%2C1\%2C1\%2C\%22X\%22\%2C\%22X\%22\%2C\%22\%E2\%97\%A6\%22\%2C\%22\%E2\%80\%A2\%22\%5D\%2C\%5B1\%2C1\%2C1\%2C1\%2C1\%2C1\%2C\%22X\%22\%2C1\%2C\%22X\%22\%2C\%22\%E2\%80\%A2\%22\%2C\%22\%E2\%80\%A2\%22\%5D\%2C\%5B1\%2C1\%2C1\%2C1\%2C1\%2C1\%2C\%22Chance3\%22\%5D\%2C\%5B\%22X\%22\%2C\%22\%E2\%80\%A2\%22\%2C1\%2C1\%2C1\%2C1\%2C\%22\%E2\%80\%A2\%22\%5D\%2C\%5B1\%2C1\%2C\%22X\%22\%2C\%22\%E2\%80\%A2\%22\%2C1\%2C1\%2C1\%2C\%22\%E2\%80\%A2\%22\%5D\%2C\%5B1\%2C1\%2C1\%2C1\%2C\%22X\%22\%2C\%22\%E2\%80\%A2\%22\%2C1\%2C1\%2C\%22\%E2\%80\%A2\%22\%5D\%2C\%5B\%22Measure\%22\%2C1\%2C\%22Measure\%22\%2C1\%2C\%22Measure\%22\%5D\%2C\%5B\%22\%E2\%80\%A2\%22\%2C\%22Z\%22\%2C1\%2C1\%2C1\%2C1\%2C\%22\%E2\%80\%A2\%22\%5D\%2C\%5B1\%2C1\%2C\%22\%E2\%80\%A2\%22\%2C\%22Z\%22\%2C1\%2C1\%2C1\%2C\%22\%E2\%80\%A2\%22\%5D\%2C\%5B1\%2C1\%2C1\%2C1\%2C\%22\%E2\%80\%A2\%22\%2C\%22Z\%22\%2C1\%2C1\%2C\%22\%E2\%80\%A2\%22\%5D\%2C\%5B1\%2C\%22H\%22\%2C1\%2C\%22H\%22\%2C1\%2C\%22H\%22\%5D\%2C\%5B1\%2C\%22Measure\%22\%2C1\%2C\%22Measure\%22\%2C1\%2C\%22Measure\%22\%5D\%2C\%5B1\%2C\%22zpar\%22\%2C1\%2C\%22zpar\%22\%2C1\%2C\%22zpar\%22\%2C1\%2C1\%2C1\%2C1\%2C1\%2C\%22X\%22\%5D\%2C\%5B1\%2C1\%2C1\%2C1\%2C1\%2C1\%2C\%22\%E2\%97\%A6\%22\%2C\%22\%E2\%97\%A6\%22\%2C\%22\%E2\%97\%A6\%22\%2C1\%2C1\%2C\%22X\%22\%5D\%2C\%5B1\%2C1\%2C1\%2C1\%2C1\%2C1\%2C1\%2C1\%2C1\%2C1\%2C1\%2C\%22Chance\%22\%5D\%5D\%2C\%22init\%22\%3A\%5B\%22\%2B\%22\%2C\%22\%2B\%22\%2C\%22\%2B\%22\%2C0\%2C\%22\%2B\%22\%2C0\%2C0\%2C0\%2C0\%2C\%22\%2B\%22\%2C\%22\%2B\%22\%5D\%7D}{Click here to open in Quirk.}
    }
    \label{fig:ghz-cz}
\end{figure}

The traditional quantum circuit for passing the GHZ test is a CSS + S circuit (see \figref{fig:ghz-s}).
We were also able to construct GHZ-passing CSS + H and CSS + CZ circuits (see \figref{fig:ghz-h} and \figref{fig:ghz-cz}).
Therefore no local hidden variable model (such as \figref{fig:sim}) can simulate CSS + H, CSS + S, or CSS + CZ circuits.
This doesn't prove there is no CSS + H simulator achieving $O(1)$ work per gate, but it presents a clear locality obstacle that would need to be overcome.

\clearpage

\section{Conclusion}
\label{sec:conclusion}

In this paper, we showed that circuits built from Pauli gates, Controlled-Not gates, X basis measurement, and Z basis measurement can be simulated in constant time per gate.
We called these circuits ``CSS circuits''.
Despite their simplicity, CSS circuits can implement a variety of quantum protocols: teleportation, superdense coding, error correction, and even entanglement purification.
However, they cannot violate Bell inequalities (unless H or S or CZ or some other gate is added to the available gates).
They can simulate some uses of entanglement, but not all uses.

An interesting practical question that we didn't explore in this paper is whether compiled CSS circuits can be efficiently reverse engineered.
For example, consider a circuit produced by individually compiling the CNOT gates of a CSS circuit into CZ+Hadamard gates.
This circuit is not strictly CSS, but it is ``secretly CSS''.
We leave the question of how to detect if a circuit is ``secretly CSS'', and how to safely and efficiently simulate such circuits, as future work.

An interesting theoretical question that we didn't explore in this paper is non-CSS-ness as a resource.
For example, note that \figref{fig:ghz-h} uses non-CSS gates to create the entangled state distributed to the players but doesn't need non-CSS gates when consuming the state.
This is an example of non-CSS-ness being encoded into a state, similar to how non-stabilizer-ness can be encoded into states like $|0\rangle + \sqrt{i} |1\rangle$~\cite{bravyi2005distillation}.

CSS circuits are such a simple concept that we cannot help but suspect we have merely rediscovered them.
For example, phase-free ZX diagrams~\cite{kissinger2022} are clearly analogous.
Regardless, we hope that readers leave this paper with fresh awareness of a useful concept.

\section{Contributions}

Cody Jones discovered CSS circuits and showed they are efficiently simulable.
Craig Gidney noticed the Bell inequality obstacle preventing the simulator from supporting other gates.
Craig Gidney wrote the paper.
No AI tools were used in the conception or writing of this paper.

\section{Acknowledgements}

We thank Hartmut Neven, and the entire Google Quantum AI team, for creating an environment where this work was possible.

\clearpage
\printbibliography

\end{document}